\documentstyle[12pt]{article}

\newcommand{\be}{\begin{equation}}
\newcommand{\ee}{\end{equation}}
\newcommand{\bn}{\begin{eqnarray}}
\newcommand{\en}{\end{eqnarray}}

\normalbaselines
\begin{document}
\begin{titlepage}
\begin{flushright}
{\bf HU-SEFT \ R \ 1996-11}
\end{flushright}
\begin{center}

\vspace*{1.0cm}

{\Large{\bf QCD Coupling Constant at Finite Temperature}
\footnote{To appear in the Festschrift dedicated to Andrzej Bialas in
 honour of his
60th birthday, Acta Physica Polonica B27 (1996).}}

\vskip 1.5cm

{\large {\bf M. Chaichian}}

\vskip 0.5cm

High Energy Physics Laboratory, Department of Physics \\ and Research
Institute for High Energy Physics, \\ University of Helsinki\\
P.O.Box 9 (Siltavuorenpenger 20 C), FIN-00014 \\ Helsinki, Finland\\

\vskip 0.2cm

$and$

\vskip 0.2cm

{\large {\bf M. Hayashi}}

\vskip 0.5cm

School of Life Science, Tokyo University of Pharmacy and Life Science,
1432-1 Horinouchi, Hachioji, Tokyo 192-03, Japan

\end{center}

\vspace{1 cm}

\begin{abstract}
\normalsize

We work out the method for evaluating the QCD coupling constant at finite
temperature ($T$) by making use of the finite $T$ renormalization group
equation up to the one-loop order on the basis of the background field
method with the imaginary time formalism. The background field method,
which maintains the explicit gauge invariance, provides notorious
simplifications since one has to calculate only the renormalization
constant of the background field gluon propagator. The results for the
evolution of the QCD coupling constant at finite $T$ reproduce partially
the ones obtained in the literature. We discuss, in particular, the origin
of the discrepancies between different calculations, such as the choice of
gauge, the break-down of Lorentz invariance, imaginary versus real time
formalism and the applicability of the Ward identities at finite $T$. 

\end{abstract}
\end{titlepage}

\renewcommand{\theequation}{\arabic{equation}}
\section{Introduction}
\setcounter{equation}{0}
\renewcommand{\theequation}{1.\arabic{equation}} 

One of the application of field theory at finite $T$ [1-4] is to find the
behaviour of coupling constant as a function of energy, temperature, and
the chemical potential using the renormalization group (RG) equation. The
knowledge of coupling constant at finite $T$ environment then can be used,
for instance, in perturbative calculations for quark-gluon plasma created
in ion-ion collisions at high energies, in the evaluation of the grand
unification scale in a cosmological context and the shift of the energy
levels in a hydrogen plasma.

A great deal of attention has been paid to the study of the behaviour 
of the QCD coupling contant at finite $T$ [5-19]. The resulting formula for
the temperature and scale dependent part of the coupling constant has
proved to be very sensitive to the prescription chosen. In this paper we
wish to reexamine this and clarify the origin of the discrepancies between
different calculations in the literature. We employ the background field
method (hereafter abbreviated as BFM)\hspace{2mm}[20,21], which is based on
a manifestly gauge invariant generating functional. The BFM provides
notorious simplifications since we have to calculate only the
renormalization constant of the background field (BF) gluon propagator. We
discuss this method in section 2.1. Since the coupling constant depends not
only on energy ($\mu$) but also on $T$, we derive a pair of RG equations -
one for $\mu$ and one for $T$. This is done in section 2.2 where we also
derive its solution.

In order to get the renormalization constant of the BF gluon propagator we
have to calculate the polarization tensor. At finite $T$ due to the lack of
the Lorentz invariance the structure of the polarization tensor is not
equivalent to the one at $T=0$. In section 3.1 we give a prescription how
to define the renormalization constant at finite $T$. In section 3.2 we
calculate the polarization tensor in the one-loop approximation. In section
3.3 calculating the vacuum part of the polarization tensor we reproduce the
well known formula for $T=0$ [22,23], which reads as 

\begin{equation}
[g^{2}(\mu)]^{-1}=[g_{0}^{2}(\mu_{0})]^{-1}
[1+2g_{0}^{2}(\mu_{0})K_{0}ln(\mu/{{\mu}_{0}})] ,
\end{equation}
where
\begin{equation}
K_{0}=(11N/3-2n_{F}/3)/(4\pi)^2.
\end{equation}
Here $N$ is for $SU(N)$, $n_{F}$ is the number of flavours, $\mu$ is the
energy scale and $\mu_0$ is the reference scale. 

We calculate the matter part of the polarization tensor in section 4.1 and,
in section 4.2, we derive the formulas for $T$ and $\mu$ dependent part of
the coupling constant.

In section 5.1 we review the results of the previous works and in section
5.2 we compare the asymptotic expansion formulas at $a=\mu/T\ll 1$ derived
in different schemes. We discuss, in particular, the origin of the
discrepancies between different calculations and give a few comments. 

\vspace{5mm}
\section{Formal methods}
\subsection{Background field method}
\setcounter{equation}{0}
\renewcommand{\theequation}{2.\arabic{equation}} 

Let us start with the generating functional for QCD:
\begin{eqnarray}
&&Z(J)=\int DA D\Psi D\bar{\Psi} D\eta D\bar{\eta}\exp\{i\int d^{4}x [{\cal
L}_{QCD}(A,\Psi,\bar{\Psi})\nonumber\\ &&-(G^{a})^{2}/(2\xi)+J^{\mu
a}A_{\mu a}+\bar{\eta} \partial G^{a}/\partial \omega^a\eta]\},
\end{eqnarray}
where $A$ is the gluon field, $\Psi$ and $\bar{\Psi}$ are the fermion
fields, $\eta$ and $\bar{\eta}$ are the ghost fields, $(G^a)^2$ is a gauge
fixing term with $\xi$ being a gauge parameter, $\omega^a$ is a $SU(3)$
group parameter, and $J^{\mu a}A_{\mu a}$ is a source term. In $SU(3)$ the
gauge fields transform as:
\begin{equation}
A_{\mu} \rightarrow U(A_{\mu}-i\partial_{\mu}/g)U^{\dagger},
\end{equation}
\begin{equation}
\Psi \rightarrow U\Psi, \bar{\Psi} \rightarrow \bar{\Psi} U^{\dagger},
\end{equation}
\begin{equation}
\eta \rightarrow U\eta, \bar{\eta} \rightarrow \bar{\eta} U^{\dagger},
\end{equation}
where $U(x)=\exp [i\omega^{a}(x)T_{a}]$ is a unitary transformation with
$T_{a}$ being a generator for $SU(3)$. By using transformations (2.2)-(2.4)
one can show that the gauge invariance of $Z(J)$ is lost in commonly used
gauges such as $\partial_\mu A_\mu=0$. Thus we see that that the explicit
gauge invariance, which is present at the classical level in gauge field
theories, is lost at the quantum level.

The advantage of the BFM [20,21] is that it can maintain the explicit gauge
invariance. For this purpose we divide the gluon field $A_\mu$ into a sum
of a classical BF $B_\mu$ and a quantum field $Q_\mu$ as
\begin{equation}
A_{\mu}=B_{\mu}+Q_{\mu},
\end{equation}
and choose for the gauge fixing term $G^{a}$ the BF gauge condition
\begin{equation}
G^{a}=(D_{\mu}(B))^{ab}Q^{\mu}_{b},
\end{equation}
where $D_{\mu}$ is the covariant derivative:
\begin{equation}
(D_{\mu})_{ab}=\partial_{\mu}\delta_{ab} +gf_{abc} B_{\mu}^{c}.
\end{equation}
Then the generating functional reads
\begin{eqnarray}
&&Z(J,B)=\int DQ D\Psi D\bar{\Psi} D\eta D\bar{\eta} \int \exp\{i\int d^{4}
x[{\cal L}_{QCD}(B+Q,\Psi,\bar{\Psi})\nonumber\\
&&-(G^{a})^{2}/(2\xi)+J^{\mu a}A_{\mu a}+\bar{\eta}\partial G^{a}/\partial
\omega^a\eta]\}.
\end{eqnarray}
This functional can be shown to be gauge invariant by using the transformations:
\begin{equation}
B_{\mu} \rightarrow U(B_{\mu}-i\partial_{\mu}/g)U^{\dagger},
\end{equation}
\begin{equation}
Q_{\mu} \rightarrow UQ_{\mu}U^{\dagger},
\end{equation}
\begin{equation}
D_{\mu}(B) \rightarrow UD_{\mu}(B)U^{\dagger},
\end{equation}
\begin{equation}
J_{\mu}(B) \rightarrow UJ_{\mu}(B)U^{\dagger},
\end{equation}
in addition to Eqs.\hspace{1mm}(2.3) and (2.4). Actually we are only
changing the integration variables in Eq. (2.8). 

In the BFM the quantum gauge fields and the ghost field need not be
renormalized since they appear only inside loops. Thus only
renormalizations
\begin{equation}
g_{0}=Z_{g}g_{R},
\end{equation}
\begin{equation}
B_{0}=(Z_{B})^{1/2}B_{R},
\end{equation}
\begin{equation}
\xi_{0} =Z_{\xi}\xi_{R},
\end{equation}
are needed.

The explicit gauge invariance of $Z(J,B)$ implies that the perturbation series 
is gauge invariant in every order in $g_R$ and that $Z_B$ and $Z_g$ are
related with each other. In order that the field tensor $F_{\mu\nu}^{a}$ in
${\cal L}_{QCD}$:
\begin{eqnarray}
F_{\mu \nu}^{a}
=(Z_{B})^{1/2}[\partial_{\mu}(B_{\nu}^{a})_{R}-\partial_{\nu}
(B^a_\mu)_R+g_RZ_g(Z_B)^{1/2}f^{abc}(B^b_{\mu})_R(B^c_{\mu})_R]
\end{eqnarray}
be gauge covariant one has to have
\begin{equation}
Z_{g}=(Z_{B})^{-1/2},
\end{equation}
which enables us to calculate the evolution of coupling constant $g_{R}$ in
a simple way. We assume this relation to be valid also at finite $T$.
\vspace{5mm}

\subsection{Coupled RG equations} 
Equations \hspace{1mm}(2.13) and (2.17) lead us to:
\begin{equation}
g_{0}=g_{R}(Z_{B})^{-1/2}.
\end{equation}
Using the dimensional regularization [24,23] let us perform our
calculations in $(4-2\varepsilon)$ dimensions. We notice from the QCD
Lagrangian that the dimension of a gluon (quark) field is
$\mu^{1-\varepsilon}(\mu^{3/2-\varepsilon}$,resp.). We redefine the bare
coupling $g_0$ so that it becomes dimensionless and rewrite Eq.(2.18) as
\begin{equation}
g_{0}=g_{R}(Z_{B})^{-1/2}\mu^{\varepsilon},
\end{equation}
where $g_{0}$ does not depend on $T$ and $\mu$. Then a pair of RG equations
results upon taking the derivative of Eq. (2.19) with respect to $T$ and
$\mu$
\begin{equation}
T\partial/\partial T[g_{R}(Z_{B})^{-1/2}\mu^{\varepsilon}]=0,
\end{equation}
\begin{equation}
\mu\partial/\partial \mu[g_{R}(Z_{B})^{-1/2}\mu^{\varepsilon}]=0. 
\end{equation}
A pair of these equations should determine the behaviour of $g_R$ with
respect to the changes in $T$ and $\mu$. Generally $Z_B$ has the form
[22,23]
\begin{equation}
Z_{B}=1+\sum_{i=1}^{\infty}g_{R}^{2i}\ [\sum_{j=1}^{i}
A_{i}^{(j)}/\varepsilon^j+f^{(i)}(\mu,T)],
\end{equation}
where $A_{j}^{(j)}/\varepsilon^{j}$ are the divergent contributions
independent on $T$ and $\mu$, while $f^{(j)}(\mu,T)$ are convergent
functions depending on $T$ and $\mu$ which vanish in the low-$T$ limit. 

Writing hereafter $g=g_R$, for simplicity, one can derive the following RG
equation from Eqs.(2.21) and (2.22) by taking the limit $\varepsilon
\rightarrow 0$:
\begin{eqnarray}
\mu\partial
g/\partial\mu=g^{3}\{-A^{(1)}+(\mu/2)\partial/\partial\mu[f^{(1)}
(\mu,T)]\}+O(g^5).
\end{eqnarray}
This equation is reduced to the ordinary one-loop RG equation in the limit
$T \rightarrow 0$ since the function $f^{(1)}(\mu,T=0)$ vanishes. 

Similarly one can derive the RG equation for $T$ from
Eqs.\hspace{1mm}(2.20) and (2.22) as:
\begin{eqnarray}
T\partial g/\partial T=(g^{3}/2)T\partial/\partial T[f^{(1)}(\mu,T)]+O(g^{5}). 
\end{eqnarray}
Equations (2.23) and (2.24) constitute the coupled RG equations. One can
easily obtain the solution to these coupled RG equations by integrating Eq.
(2.23) from $\mu_0$ to $\mu$ and Eq.(2.24) from $T_0$ to $T$. The desired
solution reads as
\begin{eqnarray}
&[g^{2}(\mu,T)]^{-1}=[g^{2}(\mu_{0},T_{0})]^{-1}+2A^{(1)}\ln(\mu/\mu_{0})
\nonumber\\
&-[f^{(1)}(\mu,T)-f^{(1)}(\mu_{0},T_{0})].
\end{eqnarray}
This equation describes the evolution of the QCD coupling constant as a
function of $T$ and $\mu$ [5-10]. $(\mu_0,T_0)$ denotes the reference
point. \vspace{5mm}
\vspace{5mm}
\section{Polarization tensor}
\subsection{Structure of the polarization tensor}
\setcounter{equation}{0}
\renewcommand{\theequation}{3.\arabic{equation}} 

At $T=0$ environment the polarization tensor is Lorentz invariant and can
be expressed as [22]
\begin{equation}
\Pi_{\mu \nu}= \Pi(k^{2}g_{\mu \nu}-k_{\mu}k_{\nu}).
\end{equation}
At $T=0$ it satisfies the transversality condition (current conservation):
\begin{equation}
k_{\mu}\Pi_{\mu \nu}=0.
\end{equation}
At finite $T$, in the presence of matter, the Lorentz invariance is lost
and the polarization tensor can only be $O(3)$ rotational invariant
[27,17]. Then it can generally depend only on 4 independent quantities,
which we can choose, for example, $\Pi_{00}, k_i\Pi_{0i}$ and the two
scalars $\Pi_L$ and $\Pi_T$ appearing in
\begin{eqnarray}
\Pi_{ij}= \Pi_{T}(\delta_{ij}-k_{i}k_{j}/\mbox{\boldmath $k$}^{2})+
\Pi_{L}k_{i}k_{j}/ \mbox{\boldmath $k$}^2.
\end{eqnarray}
At finite $T$ whether the transversality condition is satisfied or not
depends on the gauge used. In the Coulomb gauge $(\partial_iA_i=0)$, for
example, it is not transversal but is transversal at the one-loop level in
the temporal axial gauge $(A_0=0)$, and in every order of the perturbative
calculations in the BF gauge [17].

>From Eq. (3.2) we have
\begin{equation}
k_{0}\Pi_{00}=k_{i}\Pi_{i0},
\end{equation}
and from $k_{\mu}k_{\nu}\Pi_{\mu\nu}=0$, we have
\begin{equation}
k_{0}^{2}\Pi_{00}=k_{i}k_{j}\Pi_{ij}.
\end{equation}
Using Eqs.\hspace{1mm}(3.3) and (3.5) we obtain
\begin{equation}
\Pi_{L}=k_{0}^{2}\Pi_{00}/\mbox{\boldmath $k$}^{2}.
\end{equation}
For the coefficient of the transversal part of the polarization tensor one
can derive from Eqs. (3.3) and (3.6) an expression
\begin{eqnarray}
\Pi_{T}=[\Pi_{\mu \mu}-\Pi_{00}(1+k_{0}^{2}/\mbox{\boldmath $k$}^{2})]/2. 
\end{eqnarray}
Thus we have derived in Eqs.\hspace{1mm}(3.3)-(3.7) the general form of the
$T$ dependent $O(3)$ symmetric polarization tensor in the BFM.

The polarization tensor can be splitted into a sum of a $T$ independent
(vacuum) part and a $T$ and $\mu$ dependent (matter) part [4]:
\begin{equation}
\Pi_{\mu \nu}=\Pi_{\mu \nu}|_{vac}+\Pi_{\mu \nu}|_{matt} .
\end{equation}

Since the polarization tensor at $T=0$ is related to the gluon propagator,
$D_{\mu\nu}$, as
\begin{equation}
\Pi_{\mu \nu}=D_{R\mu \nu}^{-1}-D_{0 \mu \nu}^{-1}=\Pi D_{R\mu \nu}^{-1}
\end{equation}
and the relation between the bare and the renormalized propagators (see
Eq.\hspace{1mm}(2.14)) is
\begin{equation}
D_{0 \mu \nu}=Z_{B}^{-1}D_{R \mu \nu},
\end{equation}
one is led to
\begin{equation}
Z_{B}=1-\Pi.
\end{equation}
Thus the renormalization constant $Z_{B}$ can be obtained from the BF gluon
self-energy tensor $\Pi$.

In investigating the behaviour of the coupling constant at finite $T$ one
has to include the $T$ dependent parts of $\Pi$ in $Z_B$ via Eq. (3.11). In
determining $Z_{B}$ we encounter another type of ambiguity which is caused
by the lack of the Lorentz invariance. In order to define $\Pi$ at finite
$T$ we generalize Eq.(3.11) as:
\begin{equation}
Z_{B}=1-\Pi|_{vac}-\Pi|_{matt},
\end{equation}
where we have either
\begin{equation}
\Pi|_{matt}=\Pi_{00}/\mbox{\boldmath $k$}^{2},
\end{equation}
or
\begin{equation}
\Pi|_{matt}=\Pi_{T}/\mbox{\boldmath $k$}^{2}.
\end{equation}
$\Pi_{00}$ and $\Pi_{T}$ are not connected with each other and in general
there is no a priori way to decide which one is more natural.
\vspace{5mm}

\subsection{The polarization tensor at the one-loop level} 

Our working formulas, which allow to analyze the evolution of QCD coupling
constant at finite $T$, are:\hspace{1mm}Eq.(2.22), Eq.(2.25) and
Eqs.(3.12)-(3.14). Accordingly we have to evaluate the self-energy diagrams
in Fig.1 to get $Z_B$ up to the one-loop order. The Feynman rules for the
interaction vertices are the same as in the $T=0$ case, and therefore
identical to those given in [21]. Evaluating the one-loop diagrams
(1a)-(1d) in the Feynman gauge, we find for the boson contributions in the
polarization tensor
\begin{eqnarray}
&&\Pi_{\mu \nu}|_{boson}=ig^{2}N\int d^{4}p/(2\pi)^{4}[4g_{\mu \nu}k^{2}+2
(k_\mu p_\nu+k_\nu p_\mu)\nonumber\\
&&+4p_{\mu}p_{\nu}-3k_{\mu}k_{\nu}-2(k+p)^{2}g_{\mu \nu}]/\{(k+p)^2p^2\}.
\end{eqnarray}
For the fermion loop, neglecting the quark masses compared to $\mu$ and
$T$, we have from the diagram (1e):
\begin{eqnarray}
&&\Pi_{\mu \nu}|_{fermion}=-4ig^{2}T_{F}n_{F}\int
d^{4}p/(2\pi)^{4}\{k_{\mu} p_{\nu}+k_{\nu}p_{\mu}\nonumber\\
&&+2p_{\mu}p_{\nu}-g_{\mu \nu}[(kp)+p^{2} ]\}/\{(k+p)^2p^2\},
\end{eqnarray}
where $T_{F}=1/2$ for $SU(3)$.

Our polarization tensors satisfy the transversality condition
\begin{equation}
k_{\mu}\Pi_{\mu \nu}|_{boson}=k_{\mu}\Pi_{\mu \nu}|_{fermion}=0.
\end{equation}

\vspace{5mm}

\subsection{Vacuum part of the polarization tensor} 

Using a standard technique and introducing the Feynman parametrization in
Eq. (3.15) we get the formula for the boson contributions:
\begin{eqnarray}
&&\Pi_{\mu \nu}|_{boson}=ig^{2}N\int d^{4}p/(2\pi)^{4}\int_{0}^{1}
dx[4g_{\mu \nu}k^2+2(1-2x)(k_\mu p_\nu+k_\nu p_\mu)\nonumber\\
&&+4p_{\mu}p_{\nu}+(-3-4x+4x^{2})k_{\mu}k_{\nu}]/[p^{2}+k^2x(1-x)]^{2}.
\end{eqnarray}
Similarly we get the formula for the fermion contributions from
Eq.\hspace{1mm} (3.16):
\begin{eqnarray}
&&\Pi_{\mu \nu}|_{fermion}=-2ig^{2}n_{F}\int d^{4}p/(2\pi)^{4}\int_{0}^{1}
dx[2(-x+x^2)k_\mu k_\nu+(1-2x)\nonumber\\
&&\times(k_{\mu}p_{\nu}+k_{\nu}p_{\mu})-g_{\mu \nu}(kp)+2p_{\mu}p_{\nu}+
g_{\mu \nu}k^2x]/[p^2+k^2x(1-x)]^2.
\end{eqnarray}
Notice that all of the integrals are ultraviolet divergent and thus have to
be regularized. For this purpose we employ the dimensional regularization
[24,23], which preserves gauge symmetries explicitly. The integrals for the
vacuum parts become Euclidean if we change $ip_0\rightarrow p_4$ and thus
can be easily evaluated. We obtain the following results:
\begin{eqnarray}
\Pi_{\mu \nu}|_{boson}=-11g^{2}N/(3\alpha)(g_{\mu
\nu}k^{2}-k_{\mu}k_{\nu})/ \varepsilon +O(1),
\end{eqnarray}
and
\begin{eqnarray}
\Pi_{\mu \nu}|_{fermion} =2g^{2}n_{F}/(3\alpha)(g_{\mu \nu}k^{2}-k_{\mu}
k_{\nu})/\varepsilon +O(1),
\end{eqnarray}
where
\begin{equation}
\alpha=(4\pi)^{2}.
\end{equation}
Combining Eqs.\hspace{1mm}(3.20) and (3.21) and taking into account
Eqs.(2.22), (3.1) and (3.11), we reproduce the standard formula for
$A^{(1)}$ defined in Eq. (2.25), i.e., Eq.(1.2) for $T=0$. We end this
subsection by emphasizing the simplicity of the calculation by the BFM in
contrast to the conventional methods, e.g., in the covariant gauge [7-9]. 

\vspace{5mm}
\section{Temperature dependent parts of the polarization tensor}
\subsection{Calculation of $\Pi_{00}$ and $\Pi{\mu \mu}$}
\setcounter{equation}{0}
\renewcommand{\theequation}{4.\arabic{equation}} 

To specify the subtraction point in calculating $\Pi_{00}$ and $\Pi_{\mu
\mu}$ we employ the static limit of zero external energy, which is commonly
used in the literature [5]. In this prescription the momentum $k$ is
specified to be space-like $k=(0,\mbox{\boldmath $k$})$ with $k^2=-\mu^2$.
Such a choice enables us to determine the static properties. Making use of
the imaginary time formalism [25,26] from Eqs. (3.15) and (3.16) we obtain
for $\Pi_{00}$
\begin{eqnarray}
\Pi_{00}&=&g^{2}T^2(N/6+{n_{F}}/12)-2g^2N{\mu}^2[4F_{B0}(a)-F_{B2}(a)]
\\
&-&2g^2n_{F}{\mu}^2[F_{F0}(a)-F_{F2}(a)],
\end{eqnarray}
with $a=\beta\mu$. Here the boson and fermion functions $F_{in}(\beta\mu)
(i=$B,F) are defined as:
\begin{equation}
F_{in}(\beta \mu)=(1/\alpha) \int_{0}^{\infty} dx x^{n} N_{i}(\mu x/2)L,
\end{equation}
with
\begin{equation}
L =\ln|(1+x)/(1-x)|.
\end{equation}
Similarly we derive an expression for $\Pi_{\mu \mu}$:
\begin{eqnarray}
\Pi_{\mu
\mu}=g^{2}T^{2}(N/3+n_{F}/6)-2g^{2}\mu^{2}[11NF_{B0}(a)+2n_{F}F_{F0}(a)].
\end{eqnarray}

\vspace{5mm}
\subsection{Temperature dependent part of the coupling constant} 

As pointed out in section 3.1 we encounter an ambiguity in determining the
renormalization constant $Z_B$ as a direct consequence of the lack of
Lorentz invariance. Here we write two formulas, one derived from $\Pi_{00}$
and the other one from $\Pi_T$. In the prescription with
$k=(0,\mbox{\boldmath $k$})$, one has from Eq.(3.7):
\begin{eqnarray}
\Pi_{T}=(\Pi_{\mu\mu}-\Pi_{00})/2.
\end{eqnarray}
Thus from Eq. (3.13) with $\Pi_{00}$ we have
\begin{eqnarray}
&&-f^{(1)}(\mu,T)=(N/6+n_{F}/12)/a^{2}-2N[4F_{B0}(a)-F_{B2}(a)]\nonumber \\
&&-2n_{F}[F_{F0}(a)-F_{F2}(a)],
\end{eqnarray}
and the other one from Eq.(3.14) with $\Pi_{T}$
\begin{eqnarray}
&&-f^{(1)}(\mu,T)=(N/12+n_{F}/24)/a^{2}-N[7F_{B0}(a)+F_{B2}(a)]\nonumber \\
&&-n_{F}[F_{F0}(a)+F_{F2}(a)].
\end{eqnarray}

\vspace{5mm}
\section{Discussion}
\subsection{Results of previous works}
\setcounter{equation}{0}
\renewcommand{\theequation}{5.\arabic{equation}} 

Gendenshtein [5] calculated the QCD coupling constant at finite $T$ in the
one-loop approximation by using the RG equation, the dimensional
regularization, and the covariant gauge with a space-like normalization
momentum $p^\mu=(0,\mu)$ and obtained
\begin{equation}
-f^{(1)}(\mu,T)=(N/3+n_{F}/6)/a^{2}.
\end{equation}

Kajantie et al. [6] studied the gauge field part of QCD with $N$ colors.
They used the $A_0=0$ gauge and defined two renormalization schemes by
writing
\begin{eqnarray}
D_{\mu \nu}^{-1}=D_{0\mu \nu}^{-1}(Z_{A}-G/p^{2})-(F-G)P^{L}_{\mu \nu},
\end{eqnarray}
where $D_{\mu \nu}$ is the gluon propagator and $F, G$ and $P^{L}_{\mu
\nu}$ come from the polarization tensor
\begin{equation}
\Pi_{\mu \nu}=FP^{L}_{\mu \nu}+GP^{T}_{\mu \nu}.
\end{equation}
In the ``magnetic prescription'' they fixed the propagator at the point
$p^{\mu}=(0,\mu)$ and had for the $T$ depending function the expansions
(without quarks)
\begin{equation}
-f^{(1)}(\mu,T)=5N/(16a),\ \mbox{\rm for}\ a\ll 1,
\end{equation}
\begin{eqnarray}
-f^{(1)}(\mu,T)=N[17/(90a^{2})+83\alpha/(6300a^{4})],\ \mbox{\rm for}\ a\gg 1.
\end{eqnarray}
In the so-called ``electric prescription'' they derived the coupling
constant from $F$ as
\begin{eqnarray}
-f^{(1)}(\mu,T)=N[1/(3a^{2})-1/(4a)-22(\ln a)/(3\alpha)],\ \mbox{\rm for}\
a\ll 1
\end{eqnarray}
\begin{eqnarray}
-f^{(1)}(\mu,T)=-N[1/(18a^{2})-11\alpha/(900a^{4})],\ \mbox{\rm for}\ a\gg 1.
\end{eqnarray}
Notice here the change of sign in the high-$T$ behaviour. They concluded
that the ``magnetic prescription'' is more natural than the ``electric''
one because the former uses the physical part of the gluon propagator.

Nakaggawa et al.[7] used the real-time formalism and studied the
scale-parameter ratios $\Lambda(a)/\Lambda$ derived from different vertices
in 4-flavour QCD. They used the covariant gauge and found that the ratios
derived from the three-gluon vertex and the gluon-ghost vertex show just
the opposite behaviour than the one derived from the gluon-quark vertex.

Fujimoto and Yamada [8] used the real-time formalism and derived the $T$
depending coupling constant from the gauge-independent Wilson loop. At
$a\ll 1$ , it reads as (without quark contributions)
\begin{eqnarray}
-f^{(1)}(\mu,T)=C[1/(3a^{2})-1/(4a)-22(\ln a)/(3\alpha)].
\end{eqnarray}
The same authors have discussed the finite $T$ RG equations [9] in the
one-loop approximation, using the real-time formalism, the covariant gauge
and the dimensional regularization. Their results are summarized as
follows. From the gluon propagator and three-gluon vertex:
\begin{eqnarray}
&&-f^{(1)}(\mu,T)=(C+T_{f})/(4a^{2})+C[-23/3F_{B0}(a)-3F_{B2}(a)-14/3G_{B0}(
a) \nonumber\\
&&+32G_{B2}(a)]+2T_{f}[F_{F0}(a)-3F_{F2}(a)+152/9G_{F0}(a)],
\end{eqnarray}
where $T_{f}=1/2$, and $C=N$ for $SU(N)$.

From the fermion propagator and fermion-gluon vertex: 
\begin{eqnarray}
-f^{(1)}(\mu,T)&=&(C+3{C_f}+2{T_f})/(12{a^2})-2(C_f+11C){F_{B0}}(a)/3
\\&-&16(C_{f}+10C)G_{B0}(a)/9-32CG_{B2}(a)+2(-5C/6+C_f/3+T_f)
\\ &\times& F_{F0}(a)-4(8C_{f}+17C)G_{F0}(a)/9+16CG_{F2}(a),
\end{eqnarray}
where $C_{f}=(N^{2}-1)/(2N)$ for $SU(N)$. 

>From the ghost propagator and ghost-gluon vertex: 
\begin{eqnarray}
&&-f^{(1)}(\mu,T)=(C+T_{f})/(12a^{2})-C[7F_{B0}(a)+F_{B2}(a)\nonumber \\ 
&&+2G_{B0}(a)]-2T_{f}[F_{F0}(a)+F_{F2}(a)].
\end{eqnarray}
The boson and fermion functions $G_{in}(a) (i=$B,F) are defined as
\begin{eqnarray}
G_{in}(a)=(2/\alpha)\int_{0}^{\infty} dx \int_{0}^{1} dy x^{n+1}N_{i}(\mu
x/2) /[x^2(y^2+3)-1].
\end{eqnarray}
The functions $G_{in}(a)(i=$B,F) do not appear in our results in Eqs. (4.6)
and (4.7), since they come from the trigluon renormalization only.

Stephens et al. [19] performed a BF one-loop calculation of gauge invariant
beta functions at finite $T$, using the retarded/advanced formalism
developped by Aurenche and Becherrawy [28] and derived:
\begin{eqnarray}
&&-f^{(1)}(\mu,T)=(N/12+n_{F}/24)/a^{2}-N[21/4F_{B0}(a)\nonumber \\ 
&&+F_{B2}(a)+7/2G_{B1}(a)]-n_{F}[F_{F0}(a)+F_{F2}(a)].
\end{eqnarray}
The numerical coefficients of the leading terms of the high-$T$ expansion
and the fermion parts are in complete agreement with our result in Eq.(4.7).
In the boson parts, however, there are some numerical discrepancies with
our result.

\vspace{5mm}

\subsection{Comparison of asymptotic expansions} 

In order to compare our results with those mentioned in section 5.1 we
derive the asymptotic expansions for $-f^{(1)}(\mu,T)$. Equations
(5.9)-(5.11) at $a=\mu/T\ll 1$ in the high-$T$ regime read as 
follows:
\begin{eqnarray}
&&-f^{(1)}(\mu,T)=\pi C/(9\sqrt{3}a^{2})-23C/(48a)+[(-22\nonumber \\
&&+\pi/3\sqrt{3})C+(18-38\pi/9\sqrt{3})T_{f}](\ln a)/(3\alpha),
\end{eqnarray}
\begin{eqnarray}
&&-f^{(1)}(\mu,T)=[(1-\pi/\sqrt{3})C+3C_{f}]/(12a^{2})-(11C+C_{f})/(24a)
\nonumber\\
&&-[(23+82\pi/9\sqrt{3})C+(4-8\pi/9\sqrt{3})C_{f}+4T_{f}](\ln a)/(3\alpha),
\end{eqnarray}
\begin{eqnarray}
-f^{(1)}(\mu,T)=-7C/(16a)-[C(22+\pi/\sqrt{3})-8T_{f}](\ln a)/(3\alpha).
\end{eqnarray}
Next we derive the asymptotic expansion for Eqs. (4.6) at $a\ll 1$.
\begin{eqnarray}
-f^{(1)}(\mu,T)=(N/3+n_{F}/6)/a^2-N/(2a)\nonumber\\
-[N(22\ell_{1}-71/3)-4n_{f}(\ell_{2}-2/3)]/(3\alpha),
\end{eqnarray}
where
\begin{equation}
\ell_{1}=\ln (a/4\pi)+\gamma,
\end{equation}
\begin{equation}
\ell_{2}=\ln (a/\pi)+\gamma.
\end{equation}
The $T^{2}$ term coincides with the one of Gendenshtein [5], i.e.,Eq.
(5.1). The gluon part in this asymptotic formula is consistent with the
result of Elze et
al. [17] derived in the BFM:
\begin{eqnarray}
-f^{(1)}(\mu,T)=N[1/(3a^{2})-1/(2a)-22(\ln a)/(3\alpha)+\dots],\ \mbox{\rm
for} \ a\ll 1.
\end{eqnarray}
This coincides with the result of Nadkharni [13], who has the terms up to
$O(1/a)$.
Next from Eq.\hspace{1mm}(4.7) we find the asymptotic formula:
\begin{equation}
-f^{(1)}(\mu,T)=-7N/(16a)-[N(22\ell_{1}-127/6)
-4n_{F}(\ell_{2}-5/12)]/(3\alpha). 
\end{equation}

The behaviour of $\Pi_{T}$ in the infrared region is known to have a form [17]
\begin{eqnarray}
\lim \Pi_{T}(0,\mbox{\boldmath $k$})/\mu^{2}=g^{2}[-cN/a+O(\ln a)].
\end{eqnarray}
The factor $c$ has been calculated in different gauges. In the covariant
$\xi$-gauge its value is [27]
\begin{equation}
c=(9+2\xi+\xi^{2})/64,
\end{equation}
whereas in the temporal axial gauge it is [6]
\begin{equation}
c=5/16.
\end{equation}
Our formula (5.21) at small $a$ behaves as Eq.\hspace{1mm}(5.22) with
\begin{equation}
c=-7/16,\ \mbox{\rm for bosons},
\end{equation}
and
\begin{equation}
c=0,\ \mbox{\rm for fermions}.
\end{equation}
These results are consistent with the results of Refs.\hspace{1mm}[9] [see
Eq. (5.17)] and [13]. The relation (5.26) was also noticed by Elze et al.
[17]. We note here that our $c$ is negative and hence no spurious pole
appears in the transversal propagator
\begin{eqnarray}
D_{Tij}(0,\mbox{\boldmath $k$})=-(\delta_{ij}-k_{i}k_{j}/\mbox{\boldmath
$k$}^{2})/[\mbox{\boldmath $k$}^{2}-\Pi_{T} (0,\mbox{\boldmath $k$})],
\end{eqnarray}
in contrast to the covariant and the temporal axial gauge cases. 

It is a known fact that the $T^{2}$ terms are gauge independent and the
gauge parameter dependence starts at $\sim T$ order [14]. Thus the fact
that the $T^2$ terms in Eqs. (5.14) and (5.15) are different from others,
e.g. Eq. (5.1), should not originate from the gauge choices.

\vspace{5mm}
\subsection{Concluding remarks}

We end the paper by giving a few comments: 

1) In sections 5.1 and 5.2 we have seen that the $T$ dependence of the QCD
coupling constant is very sensitive to the prescription chosen. This is not

a trivial issue, because all the results obtained hitherto also heavily
depend on the vertex chosen (i.e.,the trigluon, the ghost-gluon, or the
quark-gluon vertex) to satisfy the renormalization condition of the QCD
coupling constant. Furthermore in some gauges, e.g., in the Coulomb gauge,
the transversality condition (3.2) on the polarization tensor does not hold
and hence the structure of the polarization tensor in such gauges is
different from the one which satisfies the condition.

One of the reasons why we encounter different results in the literature is
that the broken Lorentz invariance has not been treated as we have done in
Eqs. (3.13) and (3.14). For example in deriving Eqs. (5.9)-(5.11) the
authors of Ref.[9] have extracted the gluon and the vertex renormalization
constants in front of Lorentz invariant structures not paying attention to
the break-down of Lorentz invariance.

Another possible source for discrepancies is that the Ward identities are
used at finite $T$ for different renormalization constants. The derivation
of the Ward identity is based on the gauge invariance and also on the
Lorentz invariance at $T=0$. Thus it is not a surprise that results from
different gauges or even (within the same gauge) from different vertices
are totally different.

The correspondence between the imaginary and real time formalisms has been
investigated in detail [7,15-16]. The choice of the formalism can also be
the source of the discrepancies under study was pointed out in these
references. 

To clarify the issue of choosing a suitable renormalization prescription,
one would need to compute the two-loop contributions to the coupling
constant at finite $T$. It was shown, in particular, in the massive $O(N)$
scalar model that the one-loop result is drastically changed by two-loop
contributions at high $T$ and in zero momentum limit [29]. Recently Elmfors
and Kobes have also argued that the inclusion of higher-loop effects is
necessary to consistently obtain the leading term in the calculation of the
thermal beta function in a hot Yang-Mills gas [30].

2) As we have seen the function $f^{(1)}(\mu,T)$ shows a power-like
$T$-dependence instead of a logarithmic fall-off as a function of $\mu$.
Thus we have
\begin{eqnarray}
&&-[f^{(1)}(\mu,T)-f^{(1)}(\mu_{0},T_{0})]\nonumber \\ &&=\sum_{n=1,2}
c_{n}[(T/\mu)^{n}-(T_{0}/\mu_{0})^{n}]+\dots.
\end{eqnarray}
However, if a relation $T=(const)\mu$ holds, where $(const) \simeq 1/3$
[6], then one would have a logarithmic $T$-dependence like in $T=0$
environment. Such a relation holds if one is dealing with the thermal
equilibrium.

We have derived Eq.\hspace{1mm}(2.25) using two essentially different RG
equations (2.23) and (2.24). Suppose we had used only one RG equation for
$T$, i.e., Eq.(2.24), then we would have
\begin{eqnarray}
&&-[f^{(1)}(\mu,T)-f^{(1)}(\mu_{0},T)]\nonumber \\ &&=\sum_{n=1,2}
c_{n}[(T/\mu)^{n}-(T/\mu_{0})^{n}]+\dots,
\end{eqnarray}
instead of Eq.(5.28). In this case we would have a power-like dependence
even with the relation $T=(const)\mu$.

We would like also to mention that all the machinary for the evolution of
the running coupling constant at finite $T$ can be analogously applied for
the case of a quantum field theory (such as QCD) at finite energy as
formulated in [31].

In conclusion, we note in accordance with the previous observations that
there is in fact no unique way to define a $T$ depending QCD coupling
constant and the issue of finding its reasonable prescription is left as a
subject of 
further investigation.

\newpage
\begin{center}
\large {\bf {REFERENCES}}
\end{center}
\begin{enumerate}
\item L. Dolan and R. Jackiw, Phys. Rev. {\bf D9}, 3320 (1974). 
\item S. Weinberg, Phys. Rev. {\bf D9}, 3357 (1974). 
\item	For a review: N.P. Landsman and Ch.G. van Weert, Phys. Rep.
{\bf 145},141 (1987).
\item	J. Kapusta, {\em Field Theory at Finite Temperature and Density},
(Cambridge University Press, Cambridge, 1988). 
\item	L.E. Gendenshtein, Sov. J. Nucl. Phys. {\bf 29}, 841 (1979).
\item	K. Kajantie and J. Kapusta, Ann. Phys. {\bf 160}, 477 (1985);K. Enqvist
and K. Kajantie, Mod. Phys. Lett. {\bf A2}, 479 (1987). 
\item	H. Nakkagawa and A. Ni\'{e}gawa, Phys. Lett. {\bf B193}, 263 (1987);
{\bf B196}, 571(E) (1987); H. Nakkagawa, A. Ni\'{e}gawa and H. Yokota, Phys.
Rev. {\bf D38}, 2566 (1988);Phys. Lett. {\bf B244}, 63 (1990); H. Nakkagawa,
H. Yokota and A. Ni\'{e}gawa, Phys. Rev. {\bf D38}, 3211 (1988).
\item	Y. Fujimoto and H. Yamada, Phys. Lett. {\bf B212}, 77 (1988).
\item	Y. Fujimoto and H. Yamada, Phys. Lett. {\bf B200}, 167 (1988);Phys.
Lett. {\bf B195}, 231 (1987).
\item   N.P. Landsman, Phys.Lett. {\bf B232},240 (1989).
\item	R. Baier, B. Pire and D. Sciff, Phys. Lett. {\bf B238}, 367 (1990).
\item	K. Enqvist and K. Kainulainen, Z. Phys. {\bf C53},87 (1992).
\item	S. Nadkharni, Phys. Lett. {\bf B232}, 362 (1989).
\item	Y. Fujimoto and H. Yamada, Z. Phys. {\bf C40}, 365 (1988).
\item	R. Kobes, Phys. Rev. {\bf D42}, 562 (1990).
\item	R. Baier, B. Pire and D. Schiff, Z. Phys. {\bf C51}, 581 (1991).
\item	H. Elze, U. Heinz, K. Kajantie and T. Toimela, Z. Phys. {\bf C37}, 305
(1988); U. Heinz, K. Kajantie and T. Toimela, Ann. Phys. {\bf 176}, 218
(1986); H. Elze, K. Kajantie and T. Toimela, Z. Phys. {\bf C37}, 601 (1988).
\item	J. Antikainen, M. Chaichian, N.R. Pantoja and J.J. Salazar, Phys. Lett.
{\bf B242}, 412 (1990).
\item	M.A. van Eijck, C.R. Stephens and Ch.G. van Weert, Mod. Phys. Lett.
{\bf A9}, 309 (1994);D. O'Connor, C.R. Stephens and F. Freire, Mod. Phys.
Lett. {\bf A8}, 1779 (1993).
\item	B.S. DeWitt, Phys. Rev. {\bf 162}, 1195 (1967).
\item	L.F. Abbott, Nucl. Phys. {\bf B185}, 189 (1981).
\item	F.J. Yndurain, {\em Quantum Chromodynamics} (Springer-Verlag, Berlin,
1983).
\item	P. Pascual and R. Tarrach, {\em QCD: Renormalization for the
Practitioner}\\(Springer-Verlag, Berlin, 1984).
\item	G.'tHooft and M. Veltman, Nucl. Phys. {\bf B44}, 189 (1972).
\item	C.W. Bernard, Phys. Rev. {\bf D9}, 3312 (1974).
\item	K. Kinslinger and M. Morley, Phys. Rev. {\bf D13}, 2771 (1976).
\item	O.K. Kalashnikov and V.V. Klimov, Sov. J. Nucl. Phys. {\bf 31}, 699
(1980); Sov. J. Nucl. Phys. {\bf 33}, 443 (1981).
\item	P. Aurenche and T. Becherrawy, Nucl. Phys. {\bf B379}, 259 (1992).
\item	K. Funakubo and M. Sakamoto, Phys. Lett. {\bf B186}, 205 (1987).
\item   P. Elmfors and R. Kobes, Phys. Rev. {\bf D51}, 774 (1995).
\item	M. Chaichian and I. Senda, Nucl.Phys. {\bf B396}, 737 (1993);
M. Chaichian, H. Satz and I. Senda, Phys. Rev. {\bf D49}, 1566 (1994). 

\end{enumerate}

\newpage
\begin{center}
\large {\bf {Figure Caption}}
\end{center}
Fig.1. Diagrams for the one-loop calculation of the BF renormalization
factor $Z_{B}$: a) Gluon loop; b) Ghost loop; c) Gluon loop from the
4-gluon vertex; d) Ghost loop from the 2-gluon 2-ghost vertex; e) Fermion
loop. Wavy lines are quantum gauge field propagators. 
Wavy lines terminating in a "B" represent external gauge particles. Solid
lines are fermion propagators and dashed lines represent ghost propagators.

\end{document}